\documentclass{jfm}
\usepackage{graphicx}
\usepackage{epstopdf, epsfig}
\usepackage{xcolor}
\usepackage{amssymb,amsmath,multirow,hhline,bm}

\def\u{\bm{u}}

\def\x{\bm{x}}

\def\derive#1#2{\frac{\partial #1}{\partial #2}}
\def\derivee#1#2{\frac{\partial^2 #1}{\partial #2^2}}

\def\neweq#1#2{\begin{equation} \label{#1}\begin{gathered} #2 \end{gathered}\end{equation}}

\def\xloop#1{\ifx\relax#1\else[#1]\expandafter\xloop\fi}

\def\lookfori#1{\ifx i#1 \expandafter\lookfors \else \expandafter\lookfori\fi}
\def\lookfors#1{\ifx s#1 \expandafter\lookforunderscore \else \expandafter\lookfori\fi}
\def\lookforunderscore#1{\ifx \_#1 \expandafter\savefirstletter \else \expandafter\lookfori\fi}
\def\savefirstletter#1{\def\letterone{#1}\expandafter\stopatnext}
\def\stopatnext#1{\ifx \_#1 \break\fi \savesecondletter#1}
\def\savesecondletter#1{\def\lettertwo{#1}\expandafter\stopatnext}

\shorttitle{Outer streaming within a two-dimensional channel}
\shortauthor{K. Pietrzyk and I. Battiato}

\title{Outer streaming within a two-dimensional channel}

\author{Kyle Pietrzyk
  \and Ilenia Battiato\corresp{\email{ibattiat@stanford.edu}}}

\affiliation{Department of Energy Resources Engineering, Stanford University,
Stanford, CA 94305, USA}

\begin{document}

\maketitle

\begin{abstract}
Acoustic streaming is the net time-averaged flow that results from the nonlinearities in an oscillating flow. Extensive research has sought to identify different physical mechanisms and types of acoustic streaming in systems of various geometries. While streaming in a channel maintains one of the simplest geometries, dimensional analysis of the governing equations reveals that multiple regimes of streaming may occur within a channel. In this study, a framework is developed for investigating and understanding the physical streaming regimes in a two-dimensional channel. By taking different limits of the dimensionless number ratios found within the framework, streaming models derived in previous works are recovered to demonstrate the different streaming regimes within a channel. The onset of fast streaming is then analyzed with the framework and nonlinear Reynolds numbers, which indicate whether the streaming is slow or fast, are found for the different physical streaming regimes. As a result, the framework provides a base for analyzing fast streaming in a channel and streaming in multi-scale systems while organizing previous streaming models into a physical spectrum for a channel geometry.
\end{abstract}


\section{Introduction}


An unintuitive result of an oscillating flow is the net, time-averaged fluid motion generally referred to as acoustic streaming. This motion is induced by Reynolds stresses that originate from dissipated acoustic energy in a fluid \citep{Lighthill1978, Riley1998, Riley2001}. Extensive efforts have sought to classify and elucidate the physics behind different observed regimes of acoustic streaming \citep{BoluriaanandMorris2003}. Such works provide a fundamental understanding of the mechanisms behind acoustic streaming and illuminate its potential use in overcoming viscous behavior at small length scales. Microfluidic devices used for micro-mixing \citep{TsengandLin2006, HuangandXie2013}, particle manipulation \citep{CollinsandKhoo2017}, and micro-pumping \citep{TovarandLee2008, HuangandNama2014} have all benefited from acoustic streaming and have found further application in various fields of engineering including biotechnology \citep{Wu2018}, thermoacoustics \citep{Swift2002ch7}, and more recently, electrochemical storage systems \citep{HuangandLiu2020}. To better engineer microfluidic devices for specific applications, acoustic streaming has been modeled theoretically using fundamental assumptions including small streaming flow velocities \citep{MoroneyandWhite1991, PhanandCoskun2015}, incompressible flows \citep{NguyenandWhite2000}, small actuation displacements \citep{NamaandHuang2014, HintermullerandJakoby2017}, and small Mach numbers \citep{NamaandHuang2016}. While they reduce complexity in solving the governing equations, these assumptions are not valid for all systems. Therefore, justification is necessary prior to their use in a particular system.


To justify the implementation of certain assumptions, dimensional analysis may be used to reveal characteristic dimensionless numbers in the governing equations \citep{Eckart1948, Wang1965, RednikovandSadhal2011, ChiniandMalecha2014}. As demonstrated in the work of \citet{Bruus2012}, these numbers provide a relative measurement of the physical effects described in an equation and allow for systems of different dimensions to be compared, assuming they are described by the same equations. In the case that the dimensionless numbers of a system have a magnitude much large or smaller than one, potential justification for neglecting certain terms in an equation may be found based on an order of magnitude argument.

For acoustic streaming outside the boundary layer, previous works have shown that the Reynolds number commonly appears with either the Mach number or Strouhal number in the dimensionless momentum and continuity equations, depending on the chosen length scale of the system \citep{Wang1965, Qi1993}. In the case of Eckart streaming, often referred to as the ``quartz wind'', a one-dimensional momentum equation and a compressible continuity equation are used to analyze the streaming induced by an acoustic beam. In this case, the acoustic wavelength is chosen as a characteristic length scale, and the Mach number appears in the governing equations. Alternatively, in Rayleigh streaming, a geometric length in the physical domain is chosen as the characteristic length scale and the Strouhal number appears in the governing equations \citep{Riley1998, Riley2001}. While these traditional models use a single length scale to describe the dynamics of an entire system, it is possible that distinct length scales are required in each coordinate direction to accurately describe a system. This is implied in the work of \citet{HamiltonandIlinskii2002, HamiltonandIlinskii2003}, where gradients in the wave propagation direction are assumed to occur on a much larger length scale than those in the direction perpendicular to the wave propagation. For a channel geometry, the ratio between these length scales provides insight on the regime of streaming occurring within the channel. Depending on the size of the length scale ratio, certain physical effects may become dominant and justify the removal of terms in the governing equations based on order of magnitude assumptions. This ultimately determines the equations that should be used to analyze streaming flows in a justified fashion. 

In addition to simplifying the governing equations for different streaming regimes, dimensional analysis could allow for a better understand of the onset of ``fast streaming'' \citep{FriendandYeo2011}. In contrast to ``slow streaming'', which occurs when the streaming velocity is much smaller than the fluid particle velocity, fast streaming is obtained as the streaming velocity reaches the same order of magnitude as the fluid particle velocity \citep{Zarembo1971}. Though both fast and slow streaming are results of nonlinear effects, they have also been referred to as ``nonlinear'' and ``linear'' streaming, respectively, in correspondence with the type of time-averaged equations used to solve for their steady velocity fields \citep{BoluriaanandMorris2003}. In an effort to analyze fast streaming, \citet{Zarembo1971} proposed a decomposition of the dependent variables into steady and oscillating components, as successive approximation approaches alone are insufficient for analyzing fast streaming flows \citep{FriendandYeo2011}. Using a similar approach, \citet{MenguyandGilbert2000} developed an analytical model for fast streaming in a cylindrical guide assuming a low Mach number, low Shear number, and high Reynolds number. Through their analysis of this model, the authors found that the magnitude of the squared ratio between the Mach number and Shear number, referred to as the nonlinear Reynolds number, can determine whether the streaming is fast or slow. This dimensionless number has been widely used to characterize the degree to which a streaming flow in a cylindrical guide is fast or slow \citep{MoreauandBailliet2008, ReytandDaru2013, ReytandBailliet2014, DaruandReyt2017}; however, nonlinear Reynolds numbers pertaining to additional physical scenarios and geometries, including the case of a two-dimensional channel, have yet to be investigated.


In previous literature, concerns regarding general misunderstandings and the disorganization of knowledge pertaining to acoustic streaming have surfaced. According to \citet{WiklundandGreen2012}, ``[a]coustic streaming is a well-known phenomenon within the acoustics community. However, due to the many forms in which it may arise, it is often misunderstood outside of the relatively small circle of researchers actively involved in its study''. As pointed out by \citet{FriendandYeo2011}, a reason for this misunderstanding could be that ``much of what has been published in the past in this area has looked at closely related problems in acoustic streaming and the propagation of acoustic waves through fluids, yet presented in such widely different ways''. Additionally, \citet{FriendandYeo2011} maintain that ``[t]he basic equations used in the analysis remain essentially those appearing from classic derivations from 40 years and more ago" and ``the steps taken in obtaining those derivations were convenient, yet not always strictly justified". The combination of diverse presentations and ambiguously justified streaming models insists that an accepted, formally-derived, and justified framework is essential to eliminate further misunderstandings of acoustic streaming. In this framework, both classical and more recent investigations of acoustic streaming should fit into a single architecture that provides easy interpretation and contextualization of results for future studies. In pursuit of this framework, as described by \citet{FriendandYeo2011}, ``[a] thorough rederivation of the equations based on a sound mathematical and physical footing, using scaling theory as needed to properly justify the removal of terms and the use of expansion methods, seems an obvious first step in properly treating acoustofluidics".

Furthermore, the framework could provide a base with which to study fast streaming using theoretical methods. Due to the strong nonlinearities and unintuitive behaviors exhibited in fast streaming flows, a call for further development of the theory in this regime has been made by previous authors:\\

\begin{quote}
``when the nonlinearities become stronger, indeed sufficient to drive coupling across many orders of magnitude in time via dispersion and diffusion, [$\cdots$] such separation approaches could become difficult to justify. However, there are currently few alternative approaches in the literature, and strongly nonlinear acoustic phenomena at small scales remain a largely unexplored area.'' \citep{FriendandYeo2011}\\
\end{quote}

\begin{quote}
``From a theoretical point of view, there is now need to go further in the development of fast streaming studies such as the one of Menguy and Gilbert, to compare this experimental behavior to theoretical predictions, especially in the near wall region.'' \citep{MoreauandBailliet2008}\\
\end{quote}

Considering the previous statements, key features of an adequate framework to study acoustic streaming should include, but not be limited to, (i) proper handling of different ranges of scales in a consistent manner, (ii) classification of different streaming regimes, including approximations for the onset of fast streaming, and (iii) identification of the equations for different regimes in a justifiable manner.


In this work, we propose a framework for analyzing outer acoustic streaming within a two-dimensional channel. Upon implementing a suitable set of scales on the governing equations, relevant dimensionless numbers are established and compared to reveal three regimes of acoustic streaming that occur within a channel. In particular, these regimes are defined by three sets of equations previously used to model acoustic streaming within a channel: the equations used in the work of \citet{HamiltonandIlinskii2002, HamiltonandIlinskii2003}, the equations used by \citet{Riley1998, Riley2001} to describe incompressible Rayleigh streaming, and the equations used by \citet{Riley1998, Riley2001} to describe Eckart streaming. By recovering the equations for each regime in the appropriate limit, the proposed framework is validated against previous results. Discussion regarding the physical identity of each regime is also provided.

We then employ the proposed framework in conjuction with the decomposition proposed by \citet{Zarembo1971} to analyze the onset of fast streaming through time-averaged governing equations. 
The magnitudes of linear and nonlinear streaming terms in the time-averaged momentum equations are compared to reveal two distinct nonlinear Reynolds numbers in the limits of the Eckart streaming regime and the streaming regime previously analyzed by \citet{HamiltonandIlinskii2002, HamiltonandIlinskii2003}. With further analysis of the nonlinear Reynolds numbers, the role of the channel width in the transition between slow and fast streaming is assessed according to the present theory. The results of this study stand as a framework to fit previous results into the physical architecture of streaming within a channel and to assist future analyses in fast streaming and streaming in multi-scale systems, including porous media where channel geometries are canonically used as pore models.

The paper is organized as follows. In section \ref{Section_2}, the problem is formulated with a description of the system and the relevant parameters (section \ref{Section_2_1}). The governing equations and equation of state are then defined (section \ref{Section_2_2}) and the non-dimensionalization is presented in terms of the system parameters (section \ref{Section_2_3}). In scaling the equations, the dimensionless model is presented (section \ref{Section_2_4}) and validated in section \ref{Section_4} through the recovery of equation sets previously used to analyze the different regimes of streaming within a channel (sections \ref{Section_4_2_1}, \ref{Section_4_1_1}, and \ref{Section_4_1_2}). Comments are then provided in section \ref{Section_4_1_3} on the physics described in the previous sections. In section \ref{Section_5}, fast streaming is analyzed by first decomposing the dependent variables of the scaled equations into their steady and oscillating parts (section \ref{Section_5_1}). Upon obtaining the time-average of the scaled model (section \ref{Section_5_2}), the magnitudes of the dimensionless coefficients in the momentum equations are compared to find two nonlinear Reynolds numbers (sections \ref{Section_5_3_1} and \ref{Section_5_3_2}). After, a short assessment of the role of the channel width in the transition between slow and fast streaming is provided (section \ref{Section_5_3_3}). Finally, a summary of the paper is given in section \ref{Section_7}.

\section{Formulation}
\label{Section_2}

\subsection{Problem Description}
\label{Section_2_1}

\begin{figure}
\centering
\includegraphics[scale=0.28]{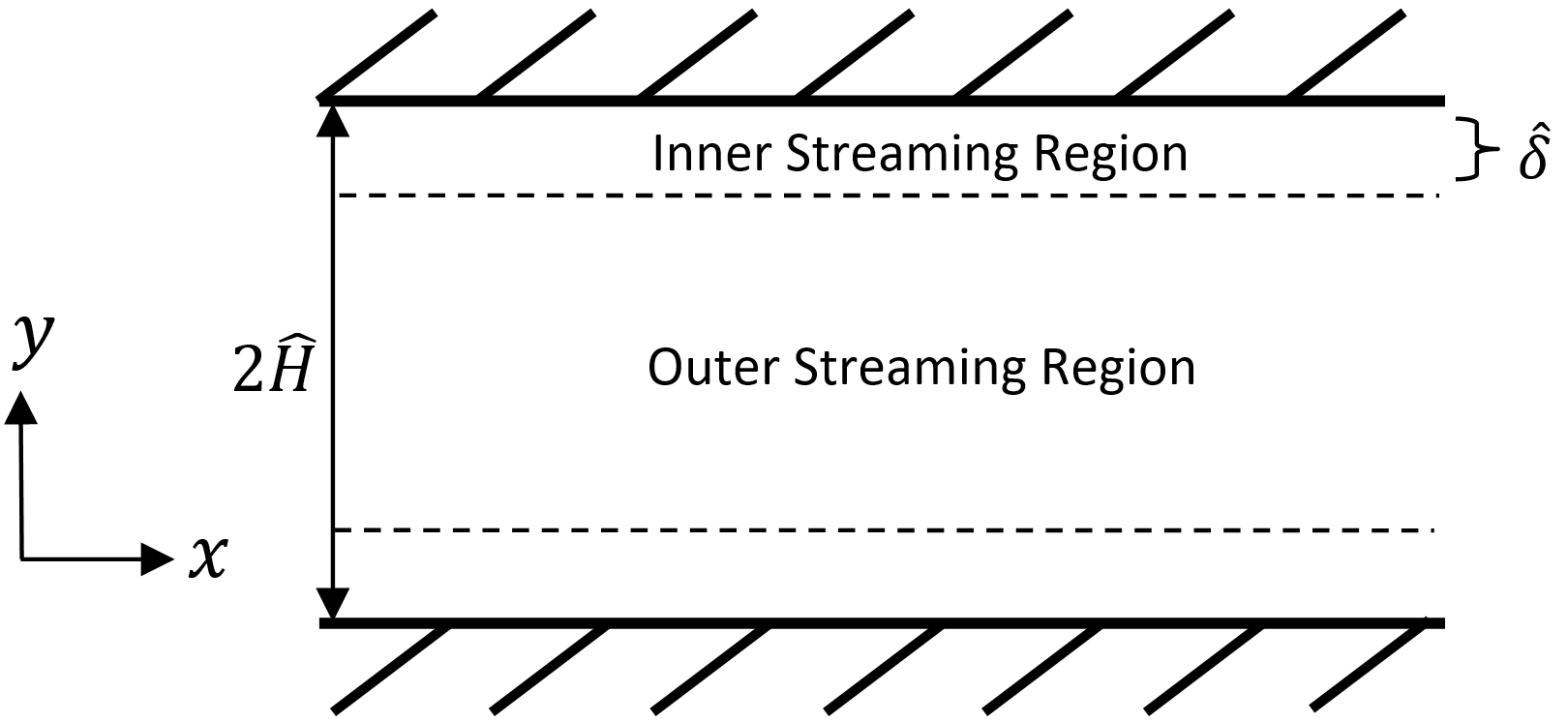}
\caption{A schematic of the geometry considered. The regions where inner and outer streaming are known to occur are labeled within a channel of width $2\hat{H}$. The height of the boundary layer is denoted by $\hat{\delta}$.}
\label{fig:1}
\end{figure}

In this study, we consider the oscillating and outer streaming flows of a Newtonian fluid in an infinitely long, two-dimensional channel of width $2\hat{H}$, as depicted in Figure \ref{fig:1}. The flow is assumed to be actuated sinusoidally at a frequency of $\hat{f} = \hat{\omega}/\left(2\pi\right)$, where $\hat{\omega}$ is the angular frequency of oscillation. The acoustic wave generated by the actuation propagates in the fluid along the channel with a velocity $\hat{c}$, the small signal speed of sound, and a wavelength of $\hat{\lambda} = \hat{c}/\hat{f}$. Only the first harmonic of the acoustic wave is considered in this analysis and no net flow is assumed to move through the channel. Lastly, the fluid is assumed to be isentropic with a bulk viscosity of $\hat{\mu}_B$, a shear viscosity of $\hat{\mu}$, and a varying density $\hat{\rho}$.

\subsection{Governing and State Equations}
\label{Section_2_2}

The motion of the fluid is governed by the Navier-Stokes equations and the mass continuity equation \citep{Batchelor1967, Zarembo1971}

\neweq{NS_eq}{\hat{\rho}\left[\derive{\hat{\u}}{\hat{t}} + \left(\hat{\u}\bm{\cdot}\hat{\nabla}\right)\hat{\u}\right] = -\hat{\nabla} \hat{p} + \hat{\mu}\hat{\nabla}^2\hat{\u} + \left(\hat{\mu}_B + \frac{\hat{\mu}}{3}\right)\hat{\nabla}\left(\hat{\nabla}\bm{\cdot}\hat{\u}\right),}

\neweq{cont_eq}{\derive{\hat{\rho}}{\hat{t}} + \hat{\nabla}\bm{\cdot}\left(\hat{\rho}\hat{\u}\right) = 0,}

\noindent where $\hat{\u}$ is the fluid velocity, $\hat{t}$ is time, and $\hat{p}$ is the pressure. Here, the hatted variables are dimensional. To close the equations, a Taylor expansion of the isentropic pressure $\hat{p} = \hat{p}\left(\hat{\rho}\right)$ about a reference density $\hat{\rho}_0$ is used with the adiabatic equation of state $\hat{c}^2 = \left.\left(\partial \hat{p}/\partial\hat{\rho}\right)\right|_{\hat{s}}$, where $\hat{s}$ is the entropy, to write \citep{NaugolnykhandOstrovsky1998, FriendandYeo2011}

\neweq{thermo_eq}{\hat{p}\left(\hat{\rho}\right) = \hat{p}\left(\hat{\rho}_0\right) + \hat{c}^2\left(\hat{\rho} - \hat{\rho}_0\right) + \frac{\hat{c}^2}{2\hat{\rho}_0}\frac{\hat{B}_0}{\hat{A}_0}\left(\hat{\rho} - \hat{\rho}_0\right)^2 + ...\;,}

\noindent where

\neweq{A_eq_and_B_eq}{\hat{A}_0 \equiv \hat{\rho}_0\left.\derive{\hat{p}}{\hat{\rho}}\right|_{\hat{s},0} = \hat{\rho}_0\hat{c}^2, \quad\quad \hat{B}_0 \equiv \hat{\rho}_0^2\left.\derivee{\hat{p}}{\hat{\rho}}\right|_{\hat{s},0}.}

\noindent While the subscript $0$ refers to reference values, the subscripts $\hat{\rho}$ and $\hat{s}$ refer to values taken at constant density and at constant entropy, respectively.

\subsection{Non-Dimensionalization}
\label{Section_2_3}

With the previously defined system, we declare the following non-dimensionalizations:

\neweq{scales}{\hat{\mu}_B = \left(\hat{\mu}\right)\mu_B, \quad \hat{t} = \left(\frac{1}{\hat{f}}\right)t, \quad \hat{\rho} = \left(\hat{\rho}_{\infty}\right)\rho, \quad \hat{p} = \left(\hat{\mathbb{P}}\right)p,    \\ 
\hat{\u} = \left(\hat{U}\right)\u, \quad \hat{p}\left(\hat{\rho}_0\right) = \left(\hat{\mathcal{P}}\right)p\left(\rho_0\right), \quad \hat{\nabla} = \left(\frac{1}{\hat{\mathcal{L}}}\right)\nabla.}

\noindent Here, $\hat{\rho}_{\infty}$ is the unperturbed density of the fluid, $\hat{\mathcal{P}}$ is a characteristic pressure related to the pressure field at reference density, $\hat{\mathbb{P}}$ is the characteristic pressure scale of the system, $\hat{U}$ is the characteristic velocity scale of the system, $1/\hat{\mathcal{L}}$ is the scale of the nabla operator, and the unhatted variables are dimensionless. A few subtleties within these chosen scales are highlighted here:

\begin{enumerate}
\item It is noted that the pressure scale $\hat{\mathbb{P}}$ is not critically important to the analysis, as substituting equation \eqref{thermo_eq} into equation \eqref{NS_eq} eliminates $\hat{\mathbb{P}}$ from the equations. However, we retain this pressure scale and choose $\hat{\mathbb{P}} = \hat{\rho}_{\infty}\hat{U}\hat{c}$ for ease in simplifying the dimensionless coefficients within the scaled governing and state equations. 
\item Secondly, the velocity scale $\hat{U}$ is interpreted by considering a theoretical piston-like actuator affecting the area of interest in the infinitely long channel. This theoretical piston is assumed to oscillate sinusoidally in time and displace the fluid in the $x$-direction with an amplitude of $\hat{A}$, such that a function $\hat{X}\left(\hat{t}\right) = \hat{A}\sin\left(\hat{\omega}\hat{t}\right)$ could describe the displacement function of the piston. In calculating the derivative of $\hat{X}\left(\hat{t}\right)$ with respect to time, the maximum velocity achieved by the piston is found as $\hat{A}\hat{\omega}$, which is used to interpret $\hat{U}$ as $\hat{U} = \hat{A}\hat{\omega}$. 
\item Finally, for a multi-scale system, defining the length scales is often an unintuitive task, as the spatial derivatives may occur on different orders depending on the physics and system geometries involved \citep{MenguyandGilbert2000}. Here, we scale the $x$-component of the spatial gradient by $1/\hat{\lambda}$, as $\hat{\lambda}$ is the order of length over which the density, velocity, and pressure vary in the $x$-direction. In the $y$-direction, the scale of the spatial gradient is not as obvious. The structures of both oscillating and streaming flows seen in the work of \citet{HamiltonandIlinskii2003} suggest that the appropriate scale for the $y$-component of the spatial gradient depends on the length of the channel width compared to that of the viscous boundary layer, which in turn depends on the fluid viscosity, fluid density, and the oscillation frequency of the system. Here, it is sufficient to scale the spatial derivative in the $y$-direction with $1/\hat{H}_y$, where $\hat{H}_y$ is defined as the appropriate length scale in the $y$-direction. In the next section, we explore the streaming regimes for different values of $\hat{H}_y$, as compared with $\hat{\lambda}$.
With $\hat{H}_y$, we write the dimensionless nabla operator as

\neweq{nabla}{\nabla = \left[\frac{\hat{\mathcal{L}}}{\hat{\lambda}} \derive{}{x} \;,\; \frac{\hat{\mathcal{L}}}{\hat{H}_y} \derive{}{y}\right].}

\noindent Here, we see that like $\hat{\mathbb{P}}$, $\hat{\mathcal{L}}$ acts as a placeholder, as it divides out when equation \eqref{nabla} is substituted into the last equality of equation \eqref{scales}.
\end{enumerate}

\subsection{Scaled Governing and State Equations}
\label{Section_2_4}

Using the set of scales defined in Eqs. \eqref{scales} and \eqref{nabla}, we rescale  \eqref{NS_eq}, \eqref{cont_eq}, and \eqref{thermo_eq} and obtain  the following dimensionless equations:

\vskip 0.1in
\noindent\underline{$x$-Momentum Equation:}

\neweq{slow_streaming1}{\rho\left[\derive{u}{t} + \epsilon^{\xi}u\derive{u}{x} + \epsilon^{\beta}v\derive{u}{y}\right] = -\derive{p}{x}                \\
+ \epsilon^{\gamma}\left[\epsilon^{2\xi}\derivee{u}{x} + \epsilon^{2\beta}\derivee{u}{y} + \epsilon^{\xi}\left(\mu_B + \frac{1}{3}\right)\left(\epsilon^{\xi}\derivee{u}{x} + \epsilon^{\beta}\frac{\partial^2v}{\partial x\partial y}\right)\right],}

\noindent\underline{$y$-Momentum Equation:}

\neweq{slow_streaming2}{\rho\left[\derive{v}{t} + \epsilon^{\xi}u\derive{v}{x} + \epsilon^{\beta}v\derive{v}{y}\right] = -\epsilon^{-\xi + \beta}\derive{p}{y}             \\
+ \epsilon^{\gamma}\left[\epsilon^{2\xi}\derivee{v}{x} + \epsilon^{2\beta}\derivee{v}{y} + \epsilon^{\beta}\left(\mu_B + \frac{1}{3}\right)\left(\epsilon^{\xi}\frac{\partial^2 u}{\partial x\partial y} + \epsilon^{\beta}\derivee{v}{y}\right)\right],}

\noindent\underline{Continuity Equation:}

\neweq{slow_streaming3}{\derive{\rho}{t} + \epsilon^{\xi}\derive{\left(\rho u\right)}{x} + \epsilon^{\beta}\derive{\left(\rho v\right)}{y} = 0,}

\noindent\underline{Thermodynamic Equation:}

\neweq{slow_streaming4}{p\left(\rho\right) = \epsilon^{\alpha}p\left(\rho_0\right) + \epsilon^{-\xi}\left[\left(\rho - \rho_0\right) + \frac{1}{2\rho_0}\frac{\hat{B}_0}{\hat{A}_0}\left(\rho - \rho_0\right)^2\right],}

\noindent where
\begin{subequations} 
\begin{align}
M&=\dfrac{\hat{U}}{\hat{\lambda}\hat{f}}=\epsilon^{\xi} ,\\  
\dfrac{1}{St}  &= \dfrac{\hat{U}}{\hat{H}_y\hat{f}}=\epsilon^{\beta},\\  
\dfrac{1}{Re} &=\dfrac{ \hat{\mu}\hat{f}}{\hat{\rho}_{\infty}\hat{U}^2}=\epsilon^{\gamma} ,\\  & \dfrac{\hat{\mathcal{P}}}{\hat{\rho}_{\infty}\hat{U}\hat{c}}= \epsilon^{\alpha}.\label{pressure}
\end{align}
\end{subequations}
Here, $M$, $St$, and $Re$ are the Mach number, Strouhal number, and Reynolds number, respectively. We also note that $\epsilon$ is a small parameter and equation \eqref{pressure} represents a pressure ratio. In using equation \eqref{slow_streaming4} to describe the pressure, we can account for pressure contributions due to incompressible fluid mechanics in the reference pressure term (subject to knowing the corresponding values of $\hat{B}_0$ and $\hat{A}_0$ for the reference pressure field) and for weakly linear and quadratic nonlinear contributions to the pressure due to a varying density \citep{NaugolnykhandOstrovsky1998}.

In the next section, the scaled equations are validated through their recovery of equation sets previously used to analyze various forms of acoustic streaming. In particular, the recovered equation sets are those used in the work of \citet{HamiltonandIlinskii2002, HamiltonandIlinskii2003}, in the work of \citet{Riley1998, Riley2001} to describe incompressible Rayleigh streaming, and in the work of \citet{Riley1998, Riley2001} to describe Eckart streaming. Though these equation sets are known to describe different types of acoustic streaming, they have not previously been considered as streaming regimes connected through a broad spectrum of streaming physics that occurs within a channel. Here, we recover these equation sets with our scaled equations using different ratios of $\epsilon^{\xi}/\epsilon^{\beta}$ to demonstrate that each set describes a different physical regime of streaming within a channel. Such regimes correspond to different combinations of length scales at which the acoustics and fluid dynamics occur.

\section{Physical Regimes of Streaming in a Two-Dimensional Channel}
\label{Section_4}

\subsection{Recovering the Equations of Hamilton et al. $\left(\epsilon^{\xi}/\epsilon^{\beta} << 1\right)$}
\label{Section_4_2_1}

To begin, we consider the regime analyzed in the work of \citet{HamiltonandIlinskii2003} by letting $\epsilon^{\xi}/\epsilon^{\beta} << 1$ ($\hat{H}_y << \hat{\lambda}$). In doing so, we rescale the velocity using a velocity scale deduced from the slow streaming results of \citet{Rayleigh1896}, as done in the work of \citet{MenguyandGilbert2000} to analyze fast streaming in a cylindrical guide:

\neweq{rescaled_velocity}{\left(\hat{U}\right)\u = \begin{bmatrix} \left(\hat{U}\right)u & \left(\sqrt{\frac{\hat{\mu}\hat{f}}{\hat{\rho}_{\infty}\hat{c}^2}}\hat{U}\right)v \end{bmatrix} \quad\rightarrow\quad \u = \begin{bmatrix} u & \epsilon^{\xi + \frac{1}{2}\gamma}v \end{bmatrix}.}

\noindent Using these scales, we rewrite equations \eqref{slow_streaming1}, \eqref{slow_streaming2}, \eqref{slow_streaming3}, and \eqref{slow_streaming4} as

\vskip 0.1in
\noindent\underline{$x$-Momentum Equation:}

\neweq{}{\rho\left[\derive{u}{t} + \epsilon^{\xi}\left(u\derive{u}{x} + \epsilon^{\beta + \frac{1}{2}\gamma}v\derive{u}{y}\right)\right] = -\derive{p}{x}                 \\
+ \epsilon^{\gamma}\left[\epsilon^{2\xi}\derivee{u}{x} + \epsilon^{2\beta}\derivee{u}{y} + \epsilon^{2\xi}\left(\mu_B + \frac{1}{3}\right)\left(\derivee{u}{x} + \epsilon^{\beta + \frac{1}{2}\gamma}\frac{\partial^2v}{\partial x\partial y}\right)\right],}

\vskip 0.1in
\noindent\underline{$y$-Momentum Equation:}

\neweq{}{\epsilon^{\xi + \frac{1}{2}\gamma}\rho\left[\derive{v}{t} + \epsilon^{\xi}\left(u\derive{v}{x} + \epsilon^{\beta + \frac{1}{2}\gamma}v\derive{v}{y}\right)\right] = -\epsilon^{-\xi + \beta}\derive{p}{y}                 \\
+ \epsilon^{\xi + \gamma}\left[\epsilon^{2\xi + \frac{1}{2}\gamma}\derivee{v}{x} + \epsilon^{2\beta + \frac{1}{2}\gamma}\derivee{v}{y} + \epsilon^{\beta}\left(\mu_B + \frac{1}{3}\right)\left(\frac{\partial^2u}{\partial x\partial y} + \epsilon^{\beta + \frac{1}{2}\gamma}\derivee{v}{y}\right)\right],}

\vskip 0.1in
\noindent\underline{Continuity Equation:}

\neweq{}{\derive{\rho}{t} + \epsilon^{\xi}\left[\derive{\left(\rho u\right)}{x} + \epsilon^{\beta + \frac{1}{2}\gamma}\derive{\left(\rho v\right)}{y}\right] = 0,}

\vskip 0.1in
\noindent\underline{Thermodynamic Equation:}

\neweq{}{p\left(\rho\right) = \epsilon^{\alpha}p\left(\rho_0\right) + \epsilon^{-\xi}\left[\left(\rho - \rho_0\right) + \frac{1}{2\rho_0}\frac{\hat{B}_0}{\hat{A}_0}\left(\rho - \rho_0\right)^2\right].}

\noindent In considering regular perturbation series solutions in powers of the Mach number, $\epsilon^{\xi}$, we yield sets of scaled equations at the leading orders similar to those used in the work of \citet{HamiltonandIlinskii2003} without the consideration of a resonator-induced body force term. Additionally, pressure contributions associated with the reference density and the weakly quadratic nonlinear density dependency are included in the current equations.

One of the defining characteristics of this regime is the acoustics occur on a much larger length scale than the fluid dynamics, as $\epsilon^{\xi}/\epsilon^{\beta} << 1$ ($\hat{H}_y << \hat{\lambda}$). Therefore, the gradients in the $x$-direction, which are governed by the acoustics for the channel geometry, are small compared to those in the $y$-direction. Additionally, we note that $\partial^2u/\partial y^2$ becomes the dominant viscous term in the momentum equations under the assumption $\epsilon^{\xi}/\epsilon^{\beta} << 1$. As a result, the dimensionless coefficient of $\partial^2u/\partial y^2$, $\epsilon^{2\beta+\gamma} = \hat{\mu}/\left(\hat{\rho}_{\infty}\hat{H}_y^2\hat{f}\right)$, serves as one over the effective Reynolds number in this regime.

\subsection{Recovering the Equations for Rayleigh Streaming $\left(\epsilon^{\xi}/\epsilon^{\beta} \sim 1\right)$}
\label{Section_4_1_1}

As the ratio $\epsilon^{\xi}/\epsilon^{\beta}$ increases to $\epsilon^{\xi}/\epsilon^{\beta} \sim 1$ ($\hat{H}_y \sim \hat{\lambda}$), we move into a regime where the acoustics and fluid dynamics occur on a similar length scale. Here, we recover the equations used in the work of \citet{Riley1998, Riley2001} to analyze incompressible Rayleigh streaming within a channel from equations \eqref{slow_streaming1}, \eqref{slow_streaming2}, and \eqref{slow_streaming3}. Upon letting $\epsilon^{\xi} \sim \epsilon^{\beta}$, we apply the curl operator to equations \eqref{slow_streaming1} and \eqref{slow_streaming2} to gain

\neweq{pre_Rayleigh_eq_mom}{\nabla\times\left\{\rho\left[\derive{\u}{t} + \epsilon^{\xi}\left(\u\bm{\cdot}\nabla\right)\u\right]\right\} = \epsilon^{2\xi + \gamma}\left[\nabla^2\bm{\zeta} + \left(\mu_B + \frac{1}{3}\right)\nabla\times\nabla\left(\nabla\bm{\cdot}\u\right)\right],}

\neweq{pre_Rayleigh_eq_cont}{\derive{\rho}{t} + \epsilon^{\xi}\nabla\bm{\cdot}\left(\rho\u\right) = 0,}

\noindent where $\bm{\zeta} = \nabla\bm{\times}\u$ is the vorticity. In general, equations \eqref{pre_Rayleigh_eq_mom} and \eqref{pre_Rayleigh_eq_cont} take into account the effects of compressibility on the density at order $\epsilon^{\xi}$. In assuming an incompressible flow, equation \eqref{pre_Rayleigh_eq_mom} results in the vorticity equation

\neweq{Rayleigh_eq}{\derive{\bm{\zeta}}{t} + \epsilon^{\xi}\left(\u\bm{\cdot}\nabla\right)\bm{\zeta} = \epsilon^{2\xi + \gamma}\nabla^2\bm{\zeta}.}

\noindent Though it is similar in structure to the equation presented by \citet{Riley1998, Riley2001} for analyzing incompressible Rayleigh streaming, this equation is less general due to the defined length scale of the system, $\hat{\lambda}\sim\hat{H}_y$. Here, we note that unlike the regime analyzed in the work of \citet{HamiltonandIlinskii2003}, the magnitudes of the viscous terms are similar. This is shown by their common dimensionless coefficient, $\epsilon^{2\xi+\gamma} = \hat{\mu}/\left(\hat{\rho}_{\infty}\hat{\lambda}^2\hat{f}\right)$, which serves as one over the effective Reynolds number in this regime.

\subsection{Recovering the Equations for Eckart Streaming $\left(\epsilon^{\xi}/\epsilon^{\beta} >> 1\right)$}
\label{Section_4_1_2}

Finally, in considering $\epsilon^{\xi}/\epsilon^{\beta} >> 1$ ($\hat{H}_y >> \hat{\lambda}$), we move into a regime where the acoustics occur over a much smaller length scale than that of the fluid dynamics. Here, we recover equations similar to those used in the work of \citet{Riley1998, Riley2001} for analyzing Eckart streaming from equations \eqref{slow_streaming1}, \eqref{slow_streaming2}, \eqref{slow_streaming3}, and \eqref{slow_streaming4}. In making a one-dimensional flow approximation due to $\epsilon^{\xi}/\epsilon^{\beta} >> 1$, neglecting the terms multiplied by $\epsilon^{\beta}$, and neglecting the quadratic nonlinear density term in equation \eqref{slow_streaming4}, we gain

\neweq{Eckart_1}{\rho\left[\derive{u}{t} + \epsilon^{\xi}u\derive{u}{x}\right] = -\derive{p}{x} + \epsilon^{2\xi + \gamma}\left(\mu_B + \frac{4}{3}\right)\derivee{u}{x},}

\neweq{Eckart_2}{\derive{\rho}{t} + \epsilon^{\xi}\derive{\left(\rho u\right)}{x} = 0,}

\neweq{Eckart_3}{p\left(\rho\right) = \epsilon^{\alpha}p\left(\rho_0\right) + \epsilon^{-\xi}\left(\rho - \rho_0\right).}

\noindent In equation \eqref{Eckart_1}, $\epsilon^{2\xi+\gamma} = \hat{\mu}/\left(\hat{\rho}_{\infty}\hat{\lambda}^2\hat{f}\right)$ is seen as one over the effective Reynolds number for this regime. To simplify the equations further, the pressure can be eliminated from the equations by substituting equation \eqref{Eckart_3} into equation \eqref{Eckart_1}. In assuming the reference pressure and density do not depend on time or space, the time derivative of equation \eqref{Eckart_1} and the spatial derivative in the $x$-direction of equation \eqref{Eckart_2} can be calculated and used to yield a similar equation to that presented by \citet{Riley1998, Riley2001} for Eckart streaming:

\neweq{Eckart_4}{\derive{}{t}\left\{\rho\derive{u}{t}\right\} - \derivee{\left(\rho u\right)}{x} - \epsilon^{2\xi + \gamma}\left(\mu_B + \frac{4}{3}\right)\frac{\partial^3 u}{\partial x^2\partial t} = -\epsilon^{\xi}\derive{}{t}\left\{\rho u\derive{u}{x}\right\}.}

\noindent This equation is similar to the one presented by \citet{Riley1998, Riley2001} with a slight difference in the treatment of the density. In Riley's work, the density has been divided into an unperturbed component, which becomes part of the dimensionless parameters in the equation, and a small perturbation component, which appears as a variable in the equation. Here, no assumptions have been made on the density, allowing it to remain in equation \eqref{Eckart_4}. Nonetheless, equations \eqref{slow_streaming1}, \eqref{slow_streaming2}, \eqref{slow_streaming3}, and \eqref{slow_streaming4} recover equations for analyzing Eckart streaming in the regime of $\epsilon^{\xi}/\epsilon^{\beta} >> 1$.

\subsection{Concluding Remarks}
\label{Section_4_1_3}

In summary, we have demonstrated that the proposed scaled equations (\eqref{slow_streaming1}, \eqref{slow_streaming2}, \eqref{slow_streaming3}, and \eqref{slow_streaming4}) recover equation sets previously used to analyze different types of streaming for various ratios of $\epsilon^{\xi}/\epsilon^{\beta}$. This implies that the previous formulations describe physical regimes within a broad spectrum of streaming physics that may occur within a channel. In particular, the physical regimes discussed are distinguished by the relative size of the length scales at which acoustic and fluid dynamic phenomena occur. In addition to these classifications, we have illuminated a relevant Reynolds number in each regime that compares inertial and viscous components in the momentum equations. The existence of these unique Reynolds numbers implies that within any of the described regimes, further classification can be made based on whether the fluid dynamics displays more inertial or viscous behavior. Therefore, determining the fluid behavior should be considered only after determining the relative size of the length scales at which the acoustics and fluid dynamics occur. In summary, we note that the regimes discussed and equations derived are simply physical classifications of acoustic streaming and are not dependent on whether the streaming is "fast" or "slow". In the next section, nonlinear Reynolds numbers are derived for the regime studied by Hamilton \textit{et al.} and the Eckart regime to determine the onset of fast streaming.


\section{Fast Streaming}
\label{Section_5}

With the scaled equations (equations \eqref{slow_streaming1}, \eqref{slow_streaming2}, \eqref{slow_streaming3}, and \eqref{slow_streaming4}), our focus turns to fast streaming. Unlike slow streaming, fast streaming is characterized by the necessity to consider nonlinear inertial components in terms of the streaming velocity to appropriately calculate the streaming velocity field. These nonlinear components warn against using regular perturbation series solutions to analyze streaming flows with strong nonlinear components. Because a different solution approach must be used to analyze fast streaming, it is crucial to identify when slow streaming evolves into fast streaming and what parameters play a role in the evolution. For the case of streaming in a cylindrical guide at low Mach numbers, answers to these inquires are provided in the work of \citet{MenguyandGilbert2000} through a dimensionless number known as the \textit{nonlinear Reynolds number}, which is defined as

\neweq{nonlinear_Reynolds_Number}{\Rey_{nl} = \frac{M^2}{Sh^2} = \frac{\hat{U}^2}{\hat{c}^2}\frac{\hat{R}^2\hat{\omega}}{\hat{\nu}},}

\noindent where $M$ is the Mach number, $Sh$ is the shear number, $\hat{R}$ is the radius of the cylindrical guide, and $\hat{\nu} = \hat{\mu}/\hat{\rho}_{\infty}$ is the kinematic viscosity. In this analysis, nonlinear Reynolds numbers for fast streaming in a two-dimensional channel are pursued.

\subsection{Decomposition of Variables into Steady and Oscillating Components}
\label{Section_5_1}

To handle the nonlinear nature of fast streaming, we follow the solution method described in the work of \citet{Zarembo1971}, where the velocity and density are decomposed into two components: a steady, time-averaged component and an oscillating component about the time-averaged component \citep{Bertelsen1973, Riley2001, RednikovandSadhal2011}. In implementing this decomposition, the dimensional velocity and density can be written as

\neweq{u_os_u_st}{\hat{\u} = \hat{\u}_{os}\left(\hat{\x},\hat{t}\right) + \hat{\u}_{st}\left(\hat{\x}\right),}

\neweq{}{\hat{\rho} = \hat{\rho}_{os}\left(\hat{\x},\hat{t}\right) + \hat{\rho}_{st}\left(\hat{\x}\right),}

\noindent where $\hat{\x}$ is the position vector and the subscripts $os$ and $st$ indicate the oscillating and steady parts of the dependent variables, respectively. Here, it is assumed that the time average of an oscillating field over one period of oscillation $\hat{T}$ is zero, where the time-averaging operator is given as

\neweq{time_avg_op}{\left<\cdot\right> = \frac{1}{\hat{T}}\int_{\hat{T}} \left(\cdot\right) \;d\hat{t}.}

\noindent In decomposing the dimensional velocity and density in this way, distinct scales may be used to appropriately non-dimensionalize each component of each variable \citep{MenguyandGilbert2000, RednikovandSadhal2011}. While $\hat{U}$ serves as an appropriate scale for the oscillating velocity, considering the previous description in section \ref{Section_2_3}, the steady velocity typically occurs at slower speeds, which necessitates a smaller velocity scale. To find this scale, we consult the description of Lagrangian and Eulerian frames of reference by \citet{Lighthill1978} in his lecture on acoustic streaming. In his explanation, Lighthill considers an oscillating Eulerian velocity field in the $x$-direction of the form

\neweq{Lighthill_model1}{\hat{u}^* = \hat{U}^*\cos\left[\hat{\omega}^*\left(\hat{t}^* - \frac{\hat{x}^*}{\hat{c}^*}\right)\right],}

\noindent where the superscript $^*$ differentiates the values used in Lighthill's model from the values used in the present work. The first order displacement of a fluid packet, $\hat{d}^*$, subject to this velocity field about an initial position $\hat{x}_0^*$ may be found by taking the integral of the velocity field with respect to time to gain

\neweq{}{\hat{d}^* = \hat{x}_0^* + \frac{\hat{U}^*}{\hat{\omega}^*}\sin\left[\hat{\omega}^*\left(\hat{t}^* - \frac{\hat{x}_0^*}{\hat{c}^*}\right)\right].}

\noindent Now, the velocity of the fluid packet as it moves slightly off the initial position $\hat{x}_0$ to position $\hat{d}^*$ may be found by Taylor expanding the velocity of the fluid packet about its initial position $\hat{x}_0^*$ to gain

\begin{align}
\label{particlev}\hat{u}_P^* &= \hat{u}^*\left(\hat{x}_0^*\right) + \left(\hat{d}^* - \hat{x}_0^*\right)\left.\derive{\hat{u}^*}{\hat{x}^*}\right|_{\hat{x}^* = \hat{x}_0^*}                \\
\nonumber&= \hat{U}^*\cos\left[\hat{\omega}^*\left(\hat{t}^* - \frac{\hat{x}_0^*}{\hat{c}^*}\right)\right] + \frac{\hat{U}^{*2}}{\hat{c}^*}\sin^2\left[\hat{\omega}^*\left(\hat{t}^* - \frac{\hat{x}_0^*}{\hat{c}^*}\right)\right].
\end{align}

\noindent While the first term in equation \eqref{particlev} has a time-averaged velocity of zero, as it represents the Eulerian velocity field contribution to the velocity of the packet, the second term has a non-zero time-averaged velocity. Therefore, even though the packet is subject to the oscillating Eulerian field in equation \eqref{Lighthill_model1}, it experiences a steady velocity. In drawing similarities between the first and second terms in equation \eqref{particlev} and the oscillating and steady components in equation \eqref{u_os_u_st}, respectively, we use the amplitudes of the terms in equation \eqref{particlev} as scales of the oscillating and steady velocities in the current problem. Because the velocity in equation \eqref{particlev} considers the Lagrangian framework and the velocity in equation \eqref{u_os_u_st} considers the Eulerian framework, these scales are only deemed appropriate when the amplitude of the Eulerian displacement field is much smaller than the wavelength. In this limit, the second term in equation \eqref{particlev} becomes negligible, and the Eulerian and Lagrangian velocities become the same. In the current problem, this limit is assumed to be accounted for by ensuring $\hat{A} << \hat{\lambda}$, where $\hat{A}$ is recalled as the displacement amplitude of a theoretical piston that actuates the fluid. In associating $\hat{U}^*$ with the present oscillating velocity scale $\hat{U}$ and $\hat{c}^*$ with the present speed of sound $\hat{c}$, we can write

\neweq{general_scaling1}{\left(\hat{U}\right)\u = \left(\hat{U}\right)\u_{os} + \left(\frac{\hat{U}^2}{\hat{c}}\right)\u_{st} \quad\rightarrow\quad \u = \u_{os} + \epsilon^{\xi}\u_{st}.}

\noindent These scales are found to be consistent with previous works \citep{Rayleigh1896, MenguyandGilbert2000}. In addition to the velocity, \citet{Lighthill1978} mentions that the associated pressure field can be defined as the product of the Eulerian velocity field, the speed of sound, and the unperturbed density. In completing the previous analysis on the pressure experienced by the fluid packet in the Eulerian velocity field, an equation for the pressure may be obtained, and in assuming an adiabatic flow such that $\hat{\rho}^* = \hat{p}^*/\hat{c}^{*2}$, the pressure on the fluid packet may be used to write an equation for the density of the fluid packet as

\neweq{dens}{\hat{\rho}_P^* = \frac{\hat{\rho}_{\infty}^*\hat{U}^*}{\hat{c}^*}\cos\left[\hat{\omega}^*\left(\hat{t}^* - \frac{\hat{x}_0^*}{\hat{c}^*}\right)\right] + \frac{\hat{\rho}_{\infty}^*\hat{U}^{*2}}{\hat{c}^{*2}}\sin^2\left[\hat{\omega}^*\left(\hat{t}^* - \frac{\hat{x}_0^*}{\hat{c}^*}\right)\right] + \hat{\rho}_{\infty}^*.}

\noindent Using similar arguments as before, we obtain $\hat{\rho}_{\infty}\hat{U}/\hat{c}$ as the scale for the oscillating density in the present work. Also, we note that the steady density in equation \eqref{dens} has two components: one pertaining to the unperturbed density and one pertaining to the time-average of the perturbed density. By dividing the steady density term in the present work such that $\hat{\rho}_{st} = \hat{\rho}_l + \hat{\rho}_{\infty}$, where $\hat{\rho}_l$ is the excess density from the time-averaging, we can scale the terms uniquely according to equation \eqref{dens} to gain

\neweq{general_scaling2}{\left(\hat{\rho}_{\infty}\right)\rho = \left(\frac{\hat{\rho}_{\infty}\hat{U}}{\hat{c}}\right)\rho_{os} + \left(\frac{\hat{\rho}_{\infty}\hat{U}^2}{\hat{c}^2}\right)\rho_l + \hat{\rho}_{\infty} \quad\rightarrow\quad \rho = \epsilon^{\xi}\rho_{os} + \epsilon^{2\xi}\rho_l + 1.}

\noindent Because the oscillating density $\rho_{os}$ and the time-averaged density component $\rho_l$ occur at the first and second orders of the Mach number, respectively, consistency is found between this scaling and previous analyses \citep{Zarembo1971, MenguyandGilbert2000}.

\subsection{Equations for Fast Streaming}
\label{Section_5_2}

With the decomposed velocity and density fields, we can now use equations \eqref{slow_streaming1}, \eqref{slow_streaming2}, \eqref{slow_streaming3}, and \eqref{slow_streaming4} to better understand the interactions between the oscillating and steady fields in fast streaming. We note that the reference density is chosen as the unperturbed density $\hat{\rho}_{\infty}$, such that the dimensionless reference density becomes $\rho_0 = 1$. While equations \eqref{slow_streaming1}, \eqref{slow_streaming2}, \eqref{slow_streaming3}, and \eqref{slow_streaming4} in terms of the decomposed dependent variables can be found in Appendix \ref{Section_10}, the corresponding time-averaged equations are given as

\vskip 0.1in
\noindent\underline{$x$-Momentum Equation:}

\neweq{fast_streaming_avg1}{\epsilon^{2\xi}\left(\epsilon^{2\xi}\rho_l + 1\right)\left[\epsilon^{\xi}u_{st}\derive{u_{st}}{x} + \epsilon^{\beta}v_{st}\derive{u_{st}}{y}\right] - \epsilon^{\xi + \gamma}\left[\epsilon^{2\xi}\derivee{u_{st}}{x} + \epsilon^{2\beta}\derive{u_{st}}{y}\right]                  \\
- \epsilon^{2\xi + \gamma}\left(\mu_B + \frac{1}{3}\right)\left[\epsilon^{\xi}\derivee{u_{st}}{x} + \epsilon^{\beta}\frac{\partial^2 v_{st}}{\partial x \partial y}\right] + \epsilon^{2\xi}\left<\rho_{os}\left(\epsilon^{\xi}u_{os}\derive{u_{st}}{x} + \epsilon^{\beta}v_{os}\derive{u_{st}}{y}            \right.\right. \\
\left.\left. + \epsilon^{\xi}u_{st}\derive{u_{os}}{x} + \epsilon^{\beta}v_{st}\derive{u_{os}}{y}\right)\right> = -\derive{\left<p\right>}{x} - \epsilon^{\xi}\left<\rho_{os}\derive{u_{os}}{t}\right>           \\
-\left<\left(\epsilon^{\xi}\rho_{os} + \epsilon^{2\xi}\rho_l + 1\right)\left(\epsilon^{\xi}u_{os}\derive{u_{os}}{x} + \epsilon^{\beta}v_{os}\derive{u_{os}}{y}\right)\right>,}

\noindent\underline{$y$-Momentum Equation:}

\neweq{fast_streaming_avg2}{\epsilon^{2\xi}\left(\epsilon^{2\xi}\rho_l + 1\right)\left[\epsilon^{\xi}u_{st}\derive{v_{st}}{x} + \epsilon^{\beta}v_{st}\derive{v_{st}}{y}\right] - \epsilon^{\xi + \gamma}\left[\epsilon^{2\xi}\derivee{v_{st}}{x} + \epsilon^{2\beta}\derive{v_{st}}{y}\right]                  \\
- \epsilon^{\xi + \beta + \gamma}\left(\mu_B + \frac{1}{3}\right)\left[\epsilon^{\xi}\frac{\partial^2u_{st}}{\partial x\partial y} + \epsilon^{\beta}\derivee{v_{st}}{y}\right] + \epsilon^{2\xi}\left<\rho_{os}\left(\epsilon^{\xi}u_{os}\derive{v_{st}}{x} + \epsilon^{\beta}v_{os}\derive{v_{st}}{y}             \right.\right. \\
\left.\left. + \epsilon^{\xi}u_{st}\derive{v_{os}}{x} + \epsilon^{\beta}v_{st}\derive{v_{os}}{y}\right)\right> = -\epsilon^{-\xi + \beta}\derive{\left<p\right>}{y} - \epsilon^{\xi}\left<\rho_{os}\derive{v_{os}}{t}\right>           \\
-\left<\left(\epsilon^{\xi}\rho_{os} + \epsilon^{2\xi}\rho_l + 1\right)\left(\epsilon^{\xi}u_{os}\derive{v_{os}}{x} + \epsilon^{\beta}v_{os}\derive{v_{os}}{y}\right)\right>,}

\noindent\underline{Continuity Equation:}

\neweq{fast_streaming_avg3}{\epsilon^{\xi}\left[\epsilon^{\xi}\derive{u_{st}}{x} + \epsilon^{\beta}\derive{v_{st}}{y}\right] + \epsilon^{3\xi}\left[\epsilon^{\xi}\derive{\left(\rho_lu_{st}\right)}{x} + \epsilon^{\beta}\derive{\left(\rho_lv_{st}\right)}{y}\right] =                    \\
-\epsilon^{\xi}\left[\epsilon^{\xi}\left<\derive{\rho_{os}u_{os}}{x}\right> + \epsilon^{\beta}\left<\derive{\rho_{os}v_{os}}{y}\right>\right],}

\noindent\underline{Thermodynamic Equation:}

\neweq{fast_streaming_avg4}{\left<p\right> = \epsilon^{\alpha}\left<p\left(1\right)\right> + \epsilon^{\xi}\rho_l + \epsilon^{\xi}\frac{\hat{B}_0}{2\hat{A}_0}\left(\left<\rho_{os}^2\right> + \epsilon^{2\xi}\rho_l^2\right).}

\noindent As shown in these equations, the previous non-dimensionalization has extracted information regarding the magnitude of each term into dimensionless coefficients. In comparing these coefficients, the degree to which nonlinear components, in terms of steady quantities, affect the fluid motion is revealed. To reduce the complexity of our analysis, we consider these equations in two separate regimes: $\epsilon^{\xi}/\epsilon^{\beta} > 1$ and $\epsilon^{\xi}/\epsilon^{\beta} < 1$. In the regime of $\epsilon^{\xi}/\epsilon^{\beta} > 1$, it is sufficient to consider equations \eqref{fast_streaming_avg1}, \eqref{fast_streaming_avg2}, \eqref{fast_streaming_avg3}, and \eqref{fast_streaming_avg4}, but in the regime of $\epsilon^{\xi}/\epsilon^{\beta} < 1$, we look to the work of \citet{MenguyandGilbert2000} to rescale the velocity components using velocity scales deduced from the slow streaming results of \citet{Rayleigh1896}. Despite using slow streaming velocity scales, \citet{MenguyandGilbert2000} derived a fast streaming model considering a cylindrical guide that agrees with experimental results up to low values of the nonlinear Reynolds number $\Rey_{nl} \sim 2$ \citep{ReytandDaru2013}. Therefore, analogous scales for streaming in a channel are used to scale the oscillating and steady velocities for the case of $\epsilon^{\xi}/\epsilon^{\beta} < 1$. In doing so, the following dimensionless decomposed velocities are obtained:

\neweq{analygous_scaling1}{\left(\hat{U}\right)u = \left(\hat{U}\right)u_{os} + \left(\frac{\hat{U}^2}{\hat{c}}\right)u_{st} \quad\rightarrow\quad u = u_{os} + \epsilon^{\xi}u_{st},}

\neweq{analygous_scaling2}{\left(\hat{U}\right)v = \left(\sqrt{\frac{\hat{\mu}\hat{f}}{\hat{\rho}_{\infty}\hat{c}^2}}\hat{U}\right)v_{os} + \left(\frac{\hat{H}_y}{\hat{\lambda}}\frac{\hat{U}^2}{\hat{c}}\right)v_{st} \quad\rightarrow\quad v = \epsilon^{\xi + \frac{1}{2}\gamma}v_{os} + \epsilon^{2\xi - \beta}v_{st}.}

\noindent As shown, the dimensionless coefficients multiplying the oscillating velocities are consistent with those from equation \eqref{rescaled_velocity}. While the implementation of these rescaled velocities into equations \eqref{slow_streaming1}, \eqref{slow_streaming2}, \eqref{slow_streaming3}, and \eqref{slow_streaming4} can be found in Appendix \ref{Section_11}, the corresponding time-averaged equations are given as

\vskip 0.1in
\noindent\underline{$x$-Momentum Equation:}

\neweq{fast_streaming_avg1_Menguy}{\epsilon^{3\xi}\left(\epsilon^{2\xi}\rho_l + 1\right)\left[u_{st}\derive{u_{st}}{x} + v_{st}\derive{u_{st}}{y}\right] - \epsilon^{\xi + \gamma}\left[\epsilon^{2\xi}\derivee{u_{st}}{x} + \epsilon^{2\beta}\derive{u_{st}}{y}\right]                  \\
- \epsilon^{3\xi + \gamma}\left(\mu_B + \frac{1}{3}\right)\left[\derivee{u_{st}}{x} + \frac{\partial^2 v_{st}}{\partial x \partial y}\right] + \epsilon^{3\xi}\left<\rho_{os}\left(u_{os}\derive{u_{st}}{x} + \epsilon^{\beta + \frac{1}{2}\gamma}v_{os}\derive{u_{st}}{y}             \right.\right. \\
\left.\left. + u_{st}\derive{u_{os}}{x} + v_{st}\derive{u_{os}}{y}\right)\right> = -\derive{\left<p\right>}{x} - \epsilon^{\xi}\left<\rho_{os}\derive{u_{os}}{t}\right>           \\
-\epsilon^{\xi}\left<\left(\epsilon^{\xi}\rho_{os} + \epsilon^{2\xi}\rho_l + 1\right)\left(u_{os}\derive{u_{os}}{x} + \epsilon^{\beta + \frac{1}{2}\gamma}v_{os}\derive{u_{os}}{y}\right)\right>,}

\noindent\underline{$y$-Momentum Equation:}

\neweq{fast_streaming_avg2_Menguy}{\epsilon^{4\xi - \beta}\left(\epsilon^{2\xi}\rho_l + 1\right)\left[u_{st}\derive{v_{st}}{x} + v_{st}\derive{v_{st}}{y}\right] - \epsilon^{2\xi - \beta + \gamma}\left[\epsilon^{2\xi}\derivee{v_{st}}{x} + \epsilon^{2\beta}\derive{v_{st}}{y}\right]                  \\
- \epsilon^{2\xi + \beta + \gamma}\left(\mu_B + \frac{1}{3}\right)\left[\frac{\partial^2u_{st}}{\partial x\partial y} + \derivee{v_{st}}{y}\right] + \epsilon^{4\xi + \frac{1}{2}\gamma}\left<\rho_{os}\left(\epsilon^{-\beta - \frac{1}{2}\gamma}u_{os}\derive{v_{st}}{x} + v_{os}\derive{v_{st}}{y}             \right.\right. \\
\left.\left. + u_{st}\derive{v_{os}}{x} + v_{st}\derive{v_{os}}{y}\right)\right> = -\epsilon^{-\xi + \beta}\derive{\left<p\right>}{y} - \epsilon^{2\xi + \frac{1}{2}\gamma}\left<\rho_{os}\derive{v_{os}}{t}\right>           \\
-\epsilon^{2\xi + \frac{1}{2}\gamma}\left<\left(\epsilon^{\xi}\rho_{os} + \epsilon^{2\xi}\rho_l + 1\right)\left(u_{os}\derive{v_{os}}{x} + \epsilon^{\beta + \frac{1}{2}\gamma}v_{os}\derive{v_{os}}{y}\right)\right>,}

\noindent\underline{Continuity Equation:}

\neweq{fast_streaming_avg3_Menguy}{\epsilon^{2\xi}\left[\derive{u_{st}}{x} + \derive{v_{st}}{y}\right] + \epsilon^{4\xi}\left[\derive{\left(\rho_lu_{st}\right)}{x} + \derive{\left(\rho_lv_{st}\right)}{y}\right] =                    \\
-\epsilon^{2\xi}\left[\left<\derive{\rho_{os}u_{os}}{x}\right> + \epsilon^{\beta + \frac{1}{2}\gamma}\left<\derive{\rho_{os}v_{os}}{y}\right>\right],}

\noindent\underline{Thermodynamic Equation:}

\neweq{fast_streaming_avg4_Menguy}{\left<p\right> = \epsilon^{\alpha}\left<p\left(1\right)\right> + \epsilon^{\xi}\rho_l + \epsilon^{\xi}\frac{\hat{B}_0}{2\hat{A}_0}\left(\left<\rho_{os}^2\right> + \epsilon^{2\xi}\rho_l^2\right).}

\subsection{Comparison of Dimensionless Coefficients}
\label{Section_5_3}

The magnitudes of the dimensionless coefficients corresponding to the linear and nonlinear streaming velocity terms in the time-averaged momentum equations are now compared to find the dominant terms assuming $\epsilon << 1$. If the largest dimensionless coefficient in an equation corresponds to a nonlinear streaming velocity term, fast streaming is achieved. Here, the dimensionless coefficients in the regimes of $\epsilon^{\xi}/\epsilon^{\beta} < 1$ and $\epsilon^{\xi}/\epsilon^{\beta} > 1$ will be compared independently, as different velocity scales are used in the analysis of each regime. In the regime of $\epsilon^{\xi}/\epsilon^{\beta} < 1$, the dimensionless coefficients in equations \eqref{fast_streaming_avg1_Menguy} and \eqref{fast_streaming_avg2_Menguy} are compared, and in the regime of $\epsilon^{\xi}/\epsilon^{\beta} > 1$, the dimensionless coefficients in equations \eqref{fast_streaming_avg1} and \eqref{fast_streaming_avg2} are compared. The dimensionless coefficients of the considered terms from the $x$ and $y$-momentum equations have been recorded in tables \ref{tab:1} and \ref{tab:2}, respectively, for the cases of $\epsilon^{\xi}/\epsilon^{\beta} < 1$ and $\epsilon^{\xi}/\epsilon^{\beta} > 1$. The nonlinear streaming terms are presented in the left columns, while the linear streaming terms are presented in the right columns.

\begin{table}
\centering
\begin{tabular}{ c  c  c  c  c  c }
\multirow{3}{*}{Nonlinear} & \multicolumn{2}{c}{Dimensionless} & \multirow{3}{*}{Linear} & \multicolumn{2}{c}{Dimensionless} \\
\multirow{3}{*}{Terms} & \multicolumn{2}{c}{Coefficients} & \multirow{3}{*}{Terms} & \multicolumn{2}{c}{Coefficients} \\
\hhline{~--~--} 
 & $\epsilon^{\xi}/\epsilon^{\beta} > 1$ & $\epsilon^{\xi}/\epsilon^{\beta} < 1$ & & $\epsilon^{\xi}/\epsilon^{\beta} > 1$ & $\epsilon^{\xi}/\epsilon^{\beta} < 1$ \\
\hline \\ [-1.2em]
$u_{st}\derive{u_{st}}{x}$ & $\epsilon^{3\xi}$ & $\epsilon^{3\xi}$ & $\left(\mu_B + \frac{4}{3}\right)\derivee{u_{st}}{x}$ & $\epsilon^{3\xi + \gamma}$ & $\epsilon^{3\xi + \gamma}$ \\ [0.5em]
$v_{st}\derive{u_{st}}{y}$ & $\epsilon^{2\xi + \beta}$ & $\epsilon^{3\xi}$ & $\derivee{u_{st}}{y}$ & $\epsilon^{\xi + 2\beta + \gamma}$ & $\epsilon^{\xi + 2\beta + \gamma}$ \\ [0.5em]
$\rho_lu_{st}\derive{u_{st}}{x}$ & $\epsilon^{5\xi}$ & $\epsilon^{5\xi}$ & $\left(\mu_B + \frac{1}{3}\right)\frac{\partial^2 v_{st}}{\partial x \partial y}$ & $\epsilon^{2\xi + \beta + \gamma}$ & $\epsilon^{3\xi + \gamma}$ \\ [0.5em]
$\rho_lv_{st}\derive{u_{st}}{y}$ & $\epsilon^{4\xi + \beta}$ & $\epsilon^{5\xi}$ & $\left<\rho_{os}\left(u_{os}\derive{u_{st}}{x} + u_{st}\derive{u_{os}}{x}\right)\right>$ & $\epsilon^{3\xi}$ & $\epsilon^{3\xi}$ \\ [0.5em]
 & & & $\left<\rho_{os}v_{os}\derive{u_{st}}{y}\right>$ & $\epsilon^{2\xi + \beta}$ & $\epsilon^{3\xi + \beta + \frac{1}{2}\gamma}$ \\ [0.5em]
& & & $\left<\rho_{os}v_{st}\derive{u_{os}}{y}\right>$ & $\epsilon^{2\xi + \beta}$ & $\epsilon^{3\xi}$ \\ [0.5em]
\end{tabular}
\caption{Terms dependent on the streaming velocities in the $x$-momentum equation. The nonlinear terms are tabulated on the left and the linear terms are tabulated on the right.}
\label{tab:1}
\end{table}

\begin{table}
\centering
\begin{tabular}{ c  c  c  c  c  c } 
\multirow{3}{*}{Nonlinear} & \multicolumn{2}{c}{Dimensionless} & \multirow{3}{*}{Linear} & \multicolumn{2}{c}{Dimensionless} \\
\multirow{3}{*}{Terms} & \multicolumn{2}{c}{Coefficients} & \multirow{3}{*}{Terms} & \multicolumn{2}{c}{Coefficients} \\
\hhline{~--~--}
 & $\epsilon^{\xi}/\epsilon^{\beta} > 1$ & $\epsilon^{\xi}/\epsilon^{\beta} < 1$ & & $\epsilon^{\xi}/\epsilon^{\beta} > 1$ & $\epsilon^{\xi}/\epsilon^{\beta} < 1$ \\
\hline \\ [-1.2em]
$u_{st}\derive{v_{st}}{x}$ & $\epsilon^{3\xi}$ & $\epsilon^{4\xi - \beta}$ & $\derivee{v_{st}}{x}$ & $\epsilon^{3\xi + \gamma}$ & $\epsilon^{4\xi - \beta + \gamma}$ \\ [0.5em]
$v_{st}\derive{v_{st}}{y}$ & $\epsilon^{2\xi + \beta}$ & $\epsilon^{4\xi - \beta}$ & $\left(\mu_B + \frac{4}{3}\right)\derivee{v_{st}}{y}$ & $\epsilon^{\xi + 2\beta + \gamma}$ & $\epsilon^{2\xi + \beta + \gamma}$ \\ [0.5em]
$\rho_lu_{st}\derive{v_{st}}{x}$ & $\epsilon^{5\xi}$ & $\epsilon^{6\xi - \beta}$ & $\left(\mu_B + \frac{1}{3}\right)\frac{\partial^2 u_{st}}{\partial x \partial y}$ & $\epsilon^{2\xi + \beta + \gamma}$ & $\epsilon^{2\xi + \beta + \gamma}$ \\ [0.5em]
$\rho_lv_{st}\derive{v_{st}}{y}$ & $\epsilon^{4\xi + \beta}$ & $\epsilon^{6\xi - \beta}$ & $\left<\rho_{os}u_{os}\derive{v_{st}}{x}\right>$ & $\epsilon^{3\xi}$ & $\epsilon^{4\xi - \beta}$ \\ [0.5em]
 & & & $\left<\rho_{os}u_{st}\derive{v_{os}}{x}\right>$ & $\epsilon^{3\xi}$ & $\epsilon^{4\xi + \frac{1}{2}\gamma}$ \\ [0.5em]
 & & & $\left<\rho_{os}\left(v_{os}\derive{v_{st}}{y} + v_{st}\derive{v_{os}}{y}\right)\right>$ & $\epsilon^{2\xi + \beta}$ & $\epsilon^{4\xi + \frac{1}{2}\gamma}$ \\ [0.5em]
\end{tabular}
\caption{Terms dependent on the streaming velocities in the $y$-momentum equation. The nonlinear terms are tabulated on the left and the linear terms are tabulated on the right.}
\label{tab:2}
\end{table}

\subsubsection{Dominant Dimensionless Coefficients for $\epsilon^{\xi}/\epsilon^{\beta} < 1$}
\label{Section_5_3_1}

Due to the variety of dimensionless coefficients in tables \ref{tab:1} and \ref{tab:2}, it is not immediately obvious when the largest dimensionless coefficient corresponds to a nonlinear term. Therefore, we first analyze the regime of $\epsilon^{\xi}/\epsilon^{\beta} < 1$ and consider when $\xi \geq 0$. We note that two equations must be obtained to calculate $u_{st}$ and $v_{st}$ for this regime. While the continuity equation (equation \eqref{fast_streaming_avg3_Menguy}) provides one linear equation, the momentum equations (equations \eqref{fast_streaming_avg1_Menguy} and \eqref{fast_streaming_avg2_Menguy}) will provide the other equation.

As shown in the tables, the largest dimensionless coefficient of the nonlinear streaming terms is $\epsilon^{3\xi}$, and therefore, $u_{st}\partial u_{st}/\partial x$ and $v_{st}\partial u_{st}/\partial y$ are the dominant nonlinear streaming terms in this regime. In considering the linear streaming terms, we find that both $\epsilon^{\xi + 2\beta + \gamma}$ and $\epsilon^{3\xi + \beta + \frac{1}{2}\gamma}$ have the potential to be the largest dimensionless coefficient depending on the values of $\beta$ and $\gamma$, as the other dimensionless coefficients are either smaller than these, smaller than the dimensionless coefficients of the nonlinear terms, or on the order of the dimensionless coefficients of the nonlinear terms. To gain a sense for how these two dimensionless coefficients compare for different values of $\beta$ and $\gamma$, we consider the case where $\beta + \frac{1}{2}\gamma = \xi$ and find that the two dimensionless coefficients are equal to $\epsilon^{4\xi}$, which is less than the dimensionless coefficient corresponding to the dominant nonlinear terms, $\epsilon^{3\xi}$. In decreasing either $\beta$ or $\gamma$ from the state described by $\beta + \frac{1}{2}\gamma = \xi$, we can make the dimensionless coefficients of the linear streaming terms greater than dimensionless coefficient corresponding to the dominant nonlinear terms. In either case, we find that $\epsilon^{\xi + 2\beta + \gamma}$ becomes greater than $\epsilon^{3\xi + \beta + \frac{1}{2}\gamma}$. This implies that when slow streaming occurs, meaning when the dominant dimensionless coefficient of the linear streaming terms is greater than that of the nonlinear streaming terms, $\epsilon^{\xi + 2\beta + \gamma}$ is always the dominant dimensionless coefficient under the specified constraints. Therefore, $\partial^2 u_{st}/\partial y^2$ becomes the dominant linear streaming term and the ratio of $\epsilon^{3\xi}$ to $\epsilon^{\xi + 2\beta + \gamma}$ alone can predict the presence of fast streaming when $\epsilon^{\xi}/\epsilon^{\beta} < 1$. In simplifying this ratio to its dimensional parameters, we find it to be the analog of the nonlinear Reynolds number presented in the work of \citet{MenguyandGilbert2000} (equation \eqref{nonlinear_Reynolds_Number}):

\neweq{Re_nl}{\Rey_{\beta} = \frac{\epsilon^{3\xi}}{\epsilon^{\xi + 2\beta + \gamma}} = \epsilon^{2\xi - 2\beta -\gamma} = \frac{\hat{U}^2}{\hat{c}^2}\frac{\hat{\rho}_{\infty}\hat{H}_y^2\hat{f}}{\hat{\mu}}, \quad\text{for}\quad \frac{\epsilon^{\xi}}{\epsilon^{\beta}} < 1.}

\noindent When $\Rey_{\beta} << 1$, slow streaming is obtained, and when $\Rey_{\beta} \geq 1$, fast streaming is obtained. We note that the nonlinear Reynolds number found here is the product of the Mach number squared and the relevant Reynolds number from the streaming regime where $\epsilon^{\xi}/\epsilon^{\beta} << 1$ described in section \ref{Section_4_2_1}. Additionally, we decompose the nonlinear Reynolds number into its dimensional parameters in equation \eqref{Re_nl} to show the relevant physical quantities involved in the transition between slow and fast streaming for the case of $\epsilon^{\xi}/\epsilon^{\beta} < 1$.

\subsubsection{Dominant Dimensionless Coefficients for $\epsilon^{\xi}/\epsilon^{\beta} > 1$}
\label{Section_5_3_2}

We now analyze the regime of $\epsilon^{\xi}/\epsilon^{\beta} > 1$ and consider the dimensionless coefficients in equations \eqref{fast_streaming_avg1} and \eqref{fast_streaming_avg2} for $\xi \geq 0$. Upon comparing the magnitudes of the dimensionless coefficients corresponding to the nonlinear streaming terms in tables \ref{tab:1} and \ref{tab:2}, we find that $u_{st}\partial u_{st}/\partial x$ and $u_{st}\partial v_{st}/\partial x$ are the dominant nonlinear streaming terms, as $\epsilon^{3\xi}$ is the largest dimensionless coefficient of the nonlinear terms. In considering the linear streaming terms, we find that both $\epsilon^{3\xi + \gamma}$ and $\epsilon^{3\xi}$ have the potential to be the largest dimensionless coefficient depending on the value of $\gamma$. However, because $\epsilon^{3\xi}$ will always be on the order of the largest dimensionless coefficient of the nonlinear terms, it does not provide potential for the dominant dimensionless coefficient of the linear terms to be greater than that of the nonlinear terms. Therefore, $\left(\mu_B + 4/3\right)\partial^2 u_{st}/\partial x^2$ and $\partial^2 v_{st}/\partial x^2$ are the dominant linear streaming terms in this regime and the ratio of $\epsilon^{3\xi}$ to $\epsilon^{3\xi + \gamma}$ is used to predict fast streaming when $\epsilon^{\xi}/\epsilon^{\beta} > 1$. In simplifying the ratio to its dimensional parameters, we find

\neweq{Re_xi}{\Rey_{\xi} = \frac{\epsilon^{3\xi}}{\epsilon^{3\xi + \gamma}} = \epsilon^{2\xi - 2\xi -\gamma} = \frac{\hat{\rho}_{\infty}\hat{U}^2}{\hat{\mu}\hat{f}}, \quad\text{for}\quad \frac{\epsilon^{\xi}}{\epsilon^{\beta}} > 1.}

\noindent When $\Rey_{\xi} << 1$, the slow streaming approximation is valid, and when $\Rey_{\xi} \geq 1$, the nonlinear streaming term must be considered. Again, we note that the nonlinear Reynolds number found here is the product of the Mach number squared and the relevant Reynolds number from the streaming regime where $\epsilon^{\xi}/\epsilon^{\beta} >> 1$ described in section \ref{Section_4_1_2}. Additionally, we decompose the nonlinear Reynolds number to its dimensional parameters in equation \eqref{Re_xi} to show the relevant physical quantities involved in the transition between slow and fast streaming in the regime of $\epsilon^{\xi}/\epsilon^{\beta} > 1$.

\begin{figure}
\centering
\includegraphics[scale=0.7]{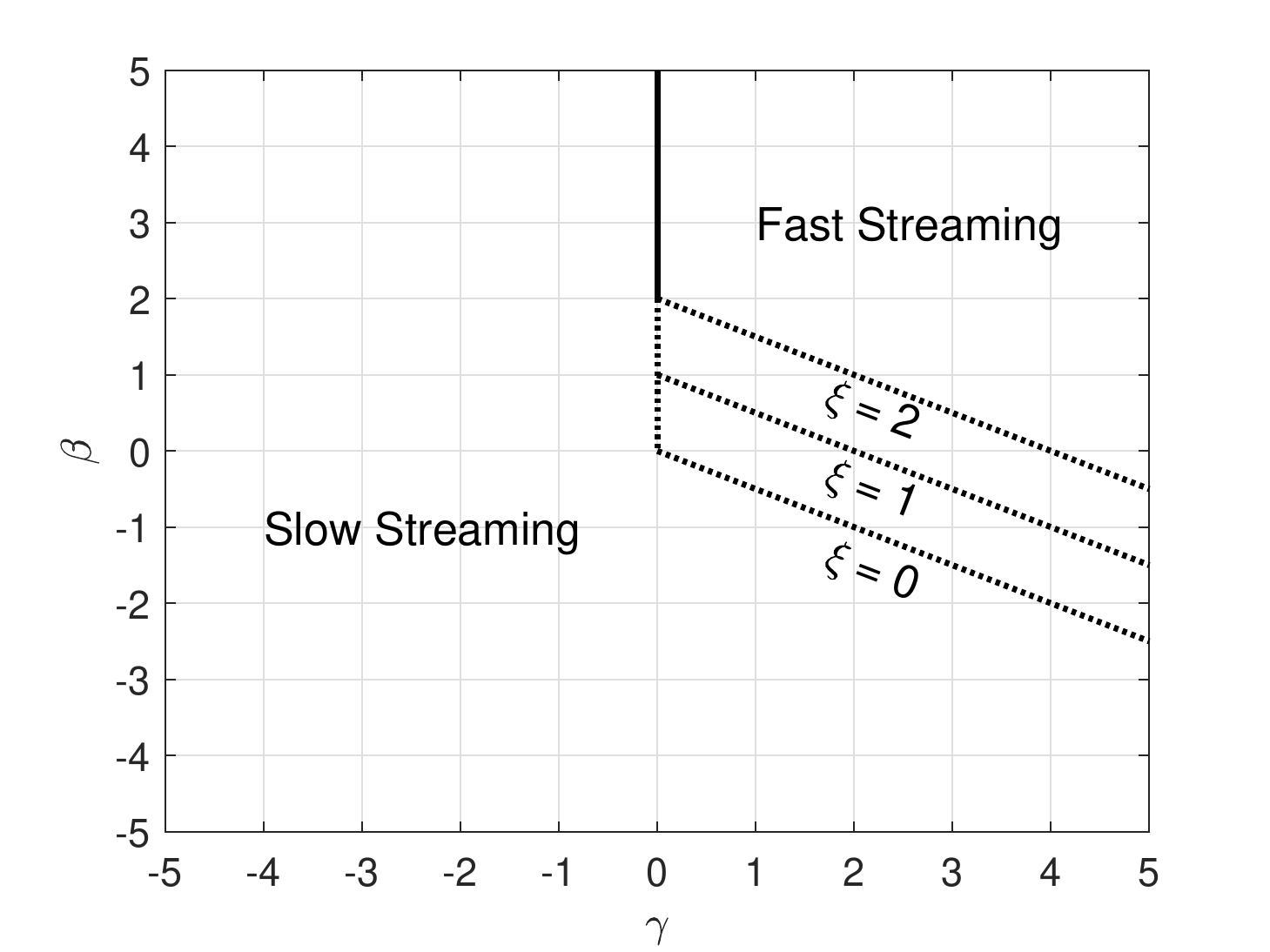}
\caption{The nonlinear Reynolds numbers, $Re_{\beta}$ and $Re_{\xi}$, plotted in the $\xi$-$\beta$-$\gamma$ phase space using $2\xi - 2\beta - \gamma = 0$ and $\gamma = 0$, respectively. Multiple cases of $Re_{\beta}$ have been plotted for different values of $\xi$. The combinations of $\xi$, $\beta$, and $\gamma$ that lead to slow and fast streaming are indicated.}
\label{fig:2}
\end{figure}

\subsubsection{Nonlinear Reynolds Number Analysis}
\label{Section_5_3_3}

Finding distinct nonlinear Reynolds numbers in the regimes of $\epsilon^{\xi}/\epsilon^{\beta} < 1$ and $\epsilon^{\xi}/\epsilon^{\beta} > 1$ sheds light on the role of the channel width in the transition between slow and fast streaming. To graphical interpret the transition between slow and fast streaming described by $Re_{\beta}$ and $Re_{\xi}$, figure \ref{fig:2} displays the $\xi-\beta-\gamma$ phase space where the exponents of $\epsilon$ from equations \eqref{Re_nl} and \eqref{Re_xi} are used to plot the divides between slow and fast streaming. From equation \eqref{Re_nl}, we plot $2\xi - 2\beta - \gamma = 0$ for $\epsilon^{\xi}/\epsilon^{\beta} < 1$, and from equation \eqref{Re_xi}, we plot $\gamma = 0$ for $\epsilon^{\xi}/\epsilon^{\beta} > 1$. 

With this depiction, we can identify two transition paths between slow and fast streaming. Considering the regime of $\epsilon^{\xi}/\epsilon^{\beta} > 1$, the first transition between slow and fast streaming occurs across the vertical divide posed by $Re_{\xi}$ at $\gamma = 0$. To cross this divide, equation \eqref{Re_xi} suggests that either the fluid properties, such as the unperturbed density $\hat{\rho}_{\infty}$ or viscosity $\hat{\mu}$, or the actuation properties, such as the frequency of actuation $\hat{f}$ or the amplitude of the actuation displacement $\hat{A}$, can be altered. In the regime of $\epsilon^{\xi}/\epsilon^{\beta} < 1$, the second transition between slow and fast streaming occurs across the slanted divide posed by $Re_{\beta}$, which can translate according to $\xi$. Like the previous divide, equation \eqref{Re_nl} suggests that the fluid properties or the actuation properties can be altered to cross the divide; however, equation \eqref{Re_nl} also suggests that $\hat{H}_y$ can be altered to cross the divide. Because the channel width influences $\hat{H}_y$, the appropriate length scale of the gradient in the $y$-direction, this theory implies the channel width can be altered to transition between slow and fast streaming. Therefore, not only do the fluid properties and actuation properties affect the transition between slow and fast streaming, but the geometry of the problem plays a role in the transition as well. The practical implication of this theory is the geometries of acoustic streaming systems may be engineered for a particular fluid and a range of actuation parameters such that the transition between slow and fast streaming happens at a known criterion and in a controlled fashion. Finally, we advise caution in using the nonlinear Reynolds numbers to predict the transition between slow and fast streaming when $\epsilon^{\xi}/\epsilon^{\beta} \sim 1$, as the magnitudes of the dimensionless coefficients in the time-averaged equations become similar in this limit.

\section{Summary}
\label{Section_7}

In this study, dimensionless equations and relevant dimensionless numbers were acquired for general outer streaming in a channel using an established set of scales. Upon varying the ratio $\epsilon^\xi/\epsilon^{\beta}$, equation sets previously used to analyze different forms of streaming were recovered. This demonstrated that these equation sets described different regimes of streaming within the physical spectrum of streaming in a channel. Such regimes were distinguished by the relative size of the length scales over which acoustic and fluid dynamic phenomena occur. Further classification of each regime could be made with the relevant Reynolds number in each regime, which shows that whether the fluid displays more inertial or viscous behavior should be considered only after comparing the acoustic and fluid dynamic length scales in the system. Also in each regime, the onset of fast streaming was defined through a unique nonlinear Reynolds number. To find the nonlinear Reynolds numbers, the dependent variables were decomposed into their oscillating and steady components and appropriate scales for each component were derived. Then, using the previously-derived dimensionless model, time-averaged governing equations were acquired and the magnitudes of the dimensionless coefficients of the linear and nonlinear streaming terms were compared in the regimes of $\epsilon^{\xi}/\epsilon^{\beta} < 1$ and $\epsilon^{\xi}/\epsilon^{\beta} > 1$. The ratios between the dimensionless coefficients of the dominant linear and nonlinear streaming terms in each regime were taken to be the nonlinear Reynolds numbers. In each regime, the nonlinear Reynolds number was found to be the product of the Mach number squared and the relevant Reynolds number of the corresponding regime. The nonlinear Reynolds numbers were then plotted in a phase diagram to illustrate the divides between slow and fast streaming. This motivated a discussion on how geometry, in addition to fluid and actuation properties, plays a role in the transition between slow and fast streaming. Overall, this study organized previous results into a framework for physically understanding streaming in a channel and derived nonlinear Reynolds numbers within regime limits. With the provided structure and physical interpretations of the defined regimes, a foundation for understanding streaming in more complex systems is gained for future analyses.

\section{Acknowledgments}
\label{Section_8}

We gratefully acknowledge the Stanford Graduate Fellowship in Science and Engineering for funding this work.

 \newpage
\appendix

\section{Fast Streaming Equations:}
\label{Section_10}

Equations \eqref{slow_streaming1}, \eqref{slow_streaming2}, \eqref{slow_streaming3}, and \eqref{slow_streaming4} with the decomposed dependent variables substituted into the equations for fast streaming analysis.

\vskip 0.1in
\noindent\underline{$x$-Momentum Equation:}

\neweq{fast_streaming1}{\left(\epsilon^{\xi}\rho_{os} + \epsilon^{2\xi}\rho_l + 1\right)\left\{\derive{u_{os}}{t} + \epsilon^{\xi}u_{os}\derive{u_{os}}{x} + \epsilon^{\beta}v_{os}\derive{u_{os}}{y} + \epsilon^{\xi}\left[\epsilon^{\xi}u_{os}\derive{u_{st}}{x}                    \right.\right. \\
\left.\left. + \epsilon^{\beta}v_{os}\derive{u_{st}}{y} + \epsilon^{\xi}u_{st}\derive{u_{os}}{x} + \epsilon^{\beta}v_{st}\derive{u_{os}}{y}\right] + \epsilon^{2\xi}\left[\epsilon^{\xi}u_{st}\derive{u_{st}}{x} + \epsilon^{\beta}v_{st}\derive{u_{st}}{y}\right]\right\} =               \\ 
-\derive{p}{x} + \epsilon^{\gamma}\left\{\epsilon^{2\xi}\derivee{u_{os}}{x} + \epsilon^{2\beta}\derivee{u_{os}}{y} + \epsilon^{\xi}\left(\mu_B + \frac{1}{3}\right)\left(\epsilon^{\xi}\derivee{u_{os}}{x} + \epsilon^{\beta}\frac{\partial^2 v_{os}}{\partial x\partial y}\right)              \right. \\
\left. + \epsilon^{\xi}\left[\epsilon^{2\xi}\derivee{u_{st}}{x} + \epsilon^{2\beta}\derivee{u_{st}}{y} + \epsilon^{\xi}\left(\mu_B + \frac{1}{3}\right)\left(\epsilon^{\xi}\derivee{u_{st}}{x} + \epsilon^{\beta}\frac{\partial^2 v_{st}}{\partial x\partial y}\right)\right]\right\},}

\noindent\underline{$y$-Momentum Equation:}

\neweq{fast_streaming2}{\left(\epsilon^{\xi}\rho_{os} + \epsilon^{2\xi}\rho_l + 1\right)\left\{\derive{v_{os}}{t} + \epsilon^{\xi}u_{os}\derive{v_{os}}{x} + \epsilon^{\beta}v_{os}\derive{v_{os}}{y} + \epsilon^{\xi}\left[\epsilon^{\xi}u_{os}\derive{v_{st}}{x}                    \right.\right. \\
\left.\left. + \epsilon^{\beta}v_{os}\derive{v_{st}}{y} + \epsilon^{\xi}u_{st}\derive{v_{os}}{x} + \epsilon^{\beta}v_{st}\derive{v_{os}}{y}\right] + \epsilon^{2\xi}\left[\epsilon^{\xi}u_{st}\derive{v_{st}}{x} + \epsilon^{\beta}v_{st}\derive{v_{st}}{y}\right]\right\} =              \\
-\epsilon^{-\xi + \beta}\derive{p}{y} + \epsilon^{\gamma}\left\{\epsilon^{2\xi}\derivee{v_{os}}{x} + \epsilon^{2\beta}\derivee{v_{os}}{y} + \epsilon^{\beta}\left(\mu_B + \frac{1}{3}\right)\left(\epsilon^{\xi}\frac{\partial^2 u_{os}}{\partial x\partial y} + \epsilon^{\beta}\derivee{v_{os}}{y}\right)         \right. \\
\left. + \epsilon^{\xi}\left[\epsilon^{2\xi}\derivee{v_{st}}{x} + \epsilon^{2\beta}\derivee{v_{st}}{y} + \epsilon^{\beta}\left(\mu_B + \frac{1}{3}\right)\left(\epsilon^{\xi}\frac{\partial^2 u_{st}}{\partial x \partial y} + \epsilon^{\beta}\derivee{v_{st}}{y}\right)\right]\right\},}

\noindent\underline{Continuity Equation:}

\neweq{fast_streaming3}{\epsilon^{\xi}\derive{\rho_{os}}{t} + \epsilon^{\xi}\derive{u_{os}}{x} + \epsilon^{\beta}\derive{v_{os}}{y}           \\
+ \epsilon^{\xi}\left[\epsilon^{\xi}\derive{u_{st}}{x} + \epsilon^{\beta}\derive{v_{st}}{y} + \epsilon^{\xi}\derive{\left(\rho_{os}u_{os}\right)}{x} + \epsilon^{\beta}\derive{\left(\rho_{os}v_{os}\right)}{y}\right]               \\
+ \epsilon^{2\xi}\left[\epsilon^{\xi}\derive{\left(\rho_{os}u_{st}\right)}{x} + \epsilon^{\beta}\derive{\left(\rho_{os}v_{st}\right)}{y} + \epsilon^{\xi}\derive{\left(\rho_lu_{os}\right)}{x} + \epsilon^{\beta}\derive{\left(\rho_lv_{os}\right)}{y}\right]            \\
+ \epsilon^{3\xi}\left[\epsilon^{\xi}\derive{\left(\rho_lu_{st}\right)}{x} + \epsilon^{\beta}\derive{\left(\rho_lv_{st}\right)}{y}\right] = 0,}

\noindent\underline{Thermodynamic Equation:}

\neweq{fast_streaming4}{p = \epsilon^{\alpha}p\left(1\right) + \left(\rho_{os} + \epsilon^{\xi}\rho_l\right) + \epsilon^{\xi}\frac{\hat{B}_0}{2\hat{A}_0}\left(\rho_{os} + \epsilon^{\xi}\rho_l\right)^2.}

\section{Fast Streaming Equations with Rescaled Velocity:}
\label{Section_11}

Equations \eqref{slow_streaming1}, \eqref{slow_streaming2}, \eqref{slow_streaming3}, and \eqref{slow_streaming4} with the decomposed dependent variables substituted into the equations for fast streaming analysis using the rescaled velocities for $\epsilon^{\xi}/\epsilon^{\beta} < 1$.

\vskip 0.1in
\noindent\underline{$x$-Momentum Equation:}

\neweq{}{\left(\epsilon^{\xi}\rho_{os} + \epsilon^{2\xi}\rho_l + 1\right)\left\{\derive{u_{os}}{t} + \epsilon^{\xi}\left[u_{os}\derive{u_{os}}{x} + \epsilon^{\beta + \frac{1}{2}\gamma}v_{os}\derive{u_{os}}{y}\right] + \epsilon^{2\xi}\left[u_{os}\derive{u_{st}}{x}                    \right.\right. \\
\left.\left. + \epsilon^{\beta + \frac{1}{2}\gamma}v_{os}\derive{u_{st}}{y} + u_{st}\derive{u_{os}}{x} + v_{st}\derive{u_{os}}{y}\right] + \epsilon^{3\xi}\left[u_{st}\derive{u_{st}}{x} + v_{st}\derive{u_{st}}{y}\right]\right\} =               \\ 
-\derive{p}{x} + \epsilon^{\gamma}\left\{\epsilon^{2\xi}\derivee{u_{os}}{x} + \epsilon^{2\beta}\derivee{u_{os}}{y} + \epsilon^{2\xi}\left(\mu_B + \frac{1}{3}\right)\left(\derivee{u_{os}}{x} + \epsilon^{\beta + \frac{1}{2}\gamma}\frac{\partial^2v_{os}}{\partial x\partial y}\right)            \right. \\
\left. + \epsilon^{\xi}\left[\epsilon^{2\xi}\derivee{u_{st}}{x} + \epsilon^{2\beta}\derivee{u_{st}}{y} + \epsilon^{2\xi}\left(\mu_B + \frac{1}{3}\right)\left(\derivee{u_{st}}{x} + \frac{\partial^2v_{st}}{\partial x\partial y}\right)\right]\right\},}

\vskip 0.1in
\noindent\underline{$y$-Momentum Equation:}

\neweq{}{\epsilon^{\xi + \frac{1}{2}\gamma}\left(\epsilon^{\xi}\rho_{os} + \epsilon^{2\xi}\rho_l + 1\right)\left\{\derive{v_{os}}{t} + \epsilon^{\xi}\left[u_{os}\derive{v_{os}}{x} + \epsilon^{\beta + \frac{1}{2}\gamma}v_{os}\derive{v_{os}}{y}\right]                    \right. \\
\left. + \epsilon^{2\xi}\left[\epsilon^{-\beta - \frac{1}{2}\gamma}u_{os}\derive{v_{st}}{x} + v_{os}\derive{v_{st}}{y} + u_{st}\derive{v_{os}}{x} + v_{st}\derive{v_{os}}{y}\right] + \epsilon^{3\xi}\left[\epsilon^{-\beta - \frac{1}{2}\gamma}u_{st}\derive{v_{st}}{x}               \right.\right. \\
\left.\left. + \epsilon^{-\beta - \frac{1}{2}\gamma}v_{st}\derive{v_{st}}{y}\right]\right\} = -\epsilon^{-\xi + \beta}\derive{p}{y} + \epsilon^{\xi + \gamma}\left\{\epsilon^{2\xi + \frac{1}{2}\gamma}\derivee{v_{os}}{x} + \epsilon^{2\beta + \frac{1}{2}\gamma}\derivee{v_{os}}{y}               \right. \\
\left. + \epsilon^{\beta}\left(\mu_B + \frac{1}{3}\right)\left(\frac{\partial^2 u_{os}}{\partial x\partial y} + \epsilon^{\beta + \frac{1}{2}\gamma}\derivee{v_{os}}{y}\right) + \epsilon^{\xi + \beta}\left[\epsilon^{2\xi - 2\beta}\derivee{v_{st}}{x} + \derivee{v_{st}}{y}              \right.\right. \\
\left.\left. + \left(\mu_B + \frac{1}{3}\right)\left(\frac{\partial^2 u_{st}}{\partial x \partial y} + \derivee{v_{st}}{y}\right)\right]\right\},}

\vskip 0.1in
\noindent\underline{Continuity Equation:}

\neweq{}{\epsilon^{\xi}\left[\derive{\rho_{os}}{t} + \derive{u_{os}}{x} + \epsilon^{\beta + \frac{1}{2}\gamma}\derive{v_{os}}{y}\right]      \\
+ \epsilon^{2\xi}\left[\derive{u_{st}}{x} + \derive{v_{st}}{y} + \derive{\left(\rho_{os}u_{os}\right)}{x} + \epsilon^{\beta + \frac{1}{2}\gamma}\derive{\left(\rho_{os}v_{os}\right)}{y}\right]        \\
+ \epsilon^{3\xi}\left[\derive{\left(\rho_{os}u_{st}\right)}{x} + \derive{\left(\rho_{os}v_{st}\right)}{y} + \derive{\left(\rho_lu_{os}\right)}{x} + \epsilon^{\beta + \frac{1}{2}\gamma}\derive{\rho_lv_{os}}{y}\right]       \\
+ \epsilon^{4\xi}\left[\derive{\left(\rho_lu_{st}\right)}{x} + \derive{\left(\rho_lv_{st}\right)}{y}\right] = 0,}

\noindent\underline{Thermodynamic Equation:}

\neweq{}{p = \epsilon^{\alpha}p\left(1\right) + \left(\rho_{os} + \epsilon^{\xi}\rho_l\right) + \epsilon^{\xi}\frac{\hat{B}_0}{2\hat{A}_0}\left(\rho_{os} + \epsilon^{\xi}\rho_l\right)^2.}

\newpage
\bibliographystyle{jfm}

\end{document}